# Memory effects of a static magnetic field on Brownian motion and the question of the absence of classical magnetism

Vladimír Lisý[1], Ján Buša[2], and Jana Tóthová[1]

[1]Faculty of Electrical Engineering and Informatics, Technical University of Košice, 04200 Košice, Slovakia
[2]Meshcheryakov Laboratory of Information Technologies, JINR, Dubna, 141980 Russia

The Bohr-van Leeuwen (BvL) theorem, stating the absence of classical magnetization in equilibrium, a fundamental result in the field of magnetic phenomena, was originally proved for an electron gas. In the present work, we deal with the problem of whether this theorem applies to particles undergoing a non-Markovian Brownian motion in a static magnetic field. We consider a charged Brownian particle (BP) immersed in a bath of neutral particles. Generalizing the Zwanzig-Caldeira-Legget theory to the presence of a static external magnetic field, we come to the equation of motion for the BP in the form of a generalized Langevin equation that accounts for memory effects in the dynamics of the system. By using its solutions for the displacement and velocity of the BP, we calculate the angular momentum for the Ornstein-Uhlenbeck thermal noise. At long times, when the system should reach equilibrium, this momentum and, consequently, the classical magnetic moment of the BP are nonzero, in contrast to the BvL theorem. With the help of analytical and precise numerical calculations for different sets of system parameters, a simple formula for the angular momentum has been deduced.

## I. INTRODUCTION

One of the basic results that can be found in any textbook on magnetic phenomenon is the Bohr-van Leeuwen (BvL) theorem [1], [2]. This theorem, in its most strong modern formulation [3], states that, in classical physics, and in thermal equilibrium, the net magnetization of a system of charged particles is zero. The magnetic field modifies the particle trajectories but does not change the total energy of the system, which is what determines the Boltzmann statistical distribution describing the thermal equilibrium. At the same time, it is understood that the BvL theorem breaks down if the system of classical charged particles uniformly rotates (as an equilibrium state) after turning on the magnetic field [4] or whenever one takes into account the internal magnetic fields produced by moving particles [5]. The theorem has also been questioned in relation to the Brownian motion (BM) of charged particles in a magnetic field [6] if described by the classical Langevin equation (LE) of motion [7]. However, the work [6] was soon criticized in [8], [9], and the existence of diamagnetism predicted in [6] was refuted. Despite the fact that now the BvL theorem is considered valid for Brownian particles (BPs) [10], [11], in the present work we returned to this problem. First of all, we were interested in whether the memory in the particle dynamics affects the mean angular momentum (and thus the magnetic moment) of BPs at long times after the application of a constant magnetic field, when the particle should be in equilibrium. We also assumed that the particle dynamics can be described by the generalized Langevin equation (GLE) [12] valid both for micrometer-sized BPs but also for atoms or molecules of the liquid itself, and the average values of the searched quantities are calculated for individual particles characterized by random initial positions and velocities of the BP and the surrounding particles of the bath in which the BP is immersed. The averaging procedure is then accomplished using the stochastic force (the colored thermal noise, not the white one as in the classical LE) included in the GLE and satisfying the fluctuation-dissipation theorem (FDT) [12]. As the basic theory, the popular Zwanzig–Caldeira–Legget (ZCL) particle-bath model has been chosen [13], [14], [15]. The equations of motion following from the ZCL model were supplemented with magnetic force acting on a charged BP, and the particle itself was surrounded by neutral particles of the liquid (bath). Although we strictly adhere to the conditions for the validity of the BvL theorem, we came to a non-zero magnetic moment of the BP at infinite time when the system should be in equilibrium, which is quite surprising as it contradicts the expectations and the cited works [10], [11].

## II. MAGNETIC MOMENT OF A CHARGED BROWNIAN PARTICLE IN NEUTRAL BATH

We consider the system of a charged Brownian particle (BP) of mass $m$ and carrying charge $Q$, immersed in a bath of $N$ particles of masses $m_i$. The system is in equilibrium up to the moment $t = 0$ when the external constant magnetic field $\mathbf{B} = (0, 0, B)$ is switched on. We restrict ourselves to the dynamics of the BP in the plane $x$, $y$ since its movement along the axis $z$ is not affected by the applied magnetic field. We come from the popular Zwanzig–Caldeira–Legget (ZCL) theory [13], [14], [15], in which the bath particles are considered oscillators with eigenfrequencies $\omega_i$. The theory, if used in the description of the Brownian motion (BM), in the absence of external fields leads to the well-known GLE [12]. The generalization of the ZCL model to the case when the BP is charged and is under the influence of an external magnetic field was considered in Refs. [16], [17], assuming that the bath particles are charged as well. Here, a simpler situation of a neutral bath will be studied. Then the equations of motion of the GLE type for the BP derived in [16] simplify to two equations

$$m\dot{\upsilon}_x(t) = QB\upsilon_y(t) - \int_0^t \upsilon_x(t')\Gamma(t-t')dt' + f_x(t), \quad (1)$$

where $\upsilon_x(t) = \dot{x}(t)$ is the projection of the velocity of the BP on the axis $x$. Here, $\Gamma(t)$ is known as the memory kernel (or memory function), and $f_x(t)$ is the projection of the random thermal force. A similar equation holds for $\upsilon_y(t) = \dot{y}(t)$ with $x \rightarrow y$ and the sign $-$ of the first term on the right-hand side of Eq. (1). The zero-mean forces $f_x(t)$ and $f_y(t)$ also differ only by changing $x$ to $y$ (when it is possible they both will be denoted as $f(t)$), they are statistically independent and determined, within the ZCL model, by the initial positions and velocities of the particles in the system,

$$f_x(t) = \sum_{i=1}^{N} m_i c_i \left\{ [x_i(0) - c_i \omega_i^{-2} x(0)] \cos(\omega_i t) + \dot{x}_i(0) \omega_i^{-1} \sin(\omega_i t) \right\}. \quad (2)$$

Here, the coefficients $c_i$ characterize the strength of coupling between the BP and the $i$th oscillator. For baths consisting of identical particles, we will below use $c_i \equiv c$ and $m_i \equiv \mu$.

Assuming the stationarity of the system, the memory function $\Gamma(t) = \sum_i m_i c_i^2 \omega_i^2 \cos(\omega t)$ is related to the thermal force by the fluctuation-dissipation theorem (FDT) [12] $\langle f(t) f(t') \rangle = k_B T \Gamma(|t - t'|)$ where $\langle ... \rangle$ means the statistical averaging. Depending on the system parameters and the distribution of frequencies $\omega_i$, $\Gamma(t)$ can be very different [14]. In what follows, we will use the exponentially decaying function $\Gamma(t) = (\xi/\tau)\exp(-t/\tau)$ corresponding to the thermal force called the Ornstein-Uhlenbeck noise, with $\xi$ interpreted as the friction coefficient of the Stokes force when the noise is white ($\Gamma(t) \sim \delta(t)$) and the integral in (1) becomes $-\xi v_x(t)$). The parameter $\tau$ has the sense of the relaxation time of the random force $f$.

We are interested in whether the ZCL theory really possesses zero magnetic moment $M_z(t)$ at $t \to \infty$ of the BP in the applied magnetic field, as it is generally assumed. For this purpose, since the magnetic moment is proportional to the angular momentum $mL(t)$, in the following we will focus on calculating the quantity $L(t) = x(t)\dot{y}(t) - y(t)\dot{x}(t)$ and its average value $\langle L(t) \rangle$ at infinite times when the system should reach the equilibrium.

To determine the positions and velocities of the BP, it is suitable to apply the Laplace transform (LT) $\mathcal{L}\{\varphi(t)\} = \tilde{\varphi}(s) = \int_0^\infty \varphi(t) e^{-st} dt$ (the inverse LT will be denoted as $\mathcal{L}^{-1}\{\tilde{\varphi}(s)\} = \varphi(t)$) to Eq. (1) and the similar equation for $v_y(t)$, which allows transforming the integro-differential equations to algebraic ones. With the rules for the operations of the LT [19], the solution of these equations can be written in the form

$$\tilde{x}(s) = \frac{x(0)}{s} + m\dot{x}(0)\tilde{\Phi}(s) + m\dot{y}(0)\tilde{\Psi}(s) + \tilde{\Phi}(s)\tilde{f}_x(s) + \tilde{\Psi}(s)\tilde{f}_y(s), \qquad (3)$$

$$\tilde{v}_x(s) = m\dot{x}(0)s\tilde{\Phi}(s) + m\dot{y}(0)s\tilde{\Psi}(s) + s\tilde{\Phi}(s)\tilde{f}_x(s) + s\tilde{\Psi}(s)\tilde{f}_y(s), \qquad (4)$$

where

$$\tilde{\Phi}(s) = \frac{ms^2 + s\tilde{\Gamma}(s)}{[ms^2 + s\tilde{\Gamma}(s)]^2 + (QBs)^2}, \qquad (5)$$

$$\tilde{\Psi}(s) = \frac{QBs}{[ms^2 + s\tilde{\Gamma}(s)]^2 + (QBs)^2}. \qquad (6)$$

with, for the chosen memory function, $\tilde{\Gamma}(s) = \xi/(\tau s + 1)$. Similar solutions take place for $\tilde{y}(s)$ and $\tilde{v}_y(s)$ if $x$ are replaced by $y$ and the terms with $\tilde{\Psi}(s)$ are written with the opposite sign. Now we have to invert these solutions and in the time domain average the result for $L(t) = x(t)v_y(t) - y(t)v_x(t)$. Taking into account Eq. (2), the independence of the different projections of the thermal force, and the fact that among the products of the initial values and velocities only $\langle x^2(0)\rangle = \langle y^2(0)\rangle$ and $\langle \dot{x}^2(0)\rangle = \langle \dot{y}^2(0)\rangle$ are nonzero, we arrive at the following terms that could contribute to $\langle L(t)\rangle$ (the terms with $x$ and $y$ contribute equally):

$A = -2\langle x(0)\mathcal{L}^{-1}\{s\tilde{\Psi}(s)\tilde{f}_x(s)\}\rangle$, $B = 2m^2\langle \dot{x}^2(0)\rangle[\mathcal{L}^{-1}\{s\tilde{\Phi}(s)\}\mathcal{L}^{-1}\{\tilde{\Psi}(s)\} - \mathcal{L}^{-1}\{\tilde{\Phi}(s)\}\mathcal{L}^{-1}\{s\tilde{\Psi}(s)\}]$,

and $C = 2[\mathcal{L}^{-1}\{s\tilde{\Phi}(s)\tilde{f}(s)\}\mathcal{L}^{-1}\{\tilde{\Psi}(s)\tilde{f}(s)\} - \mathcal{L}^{-1}\{\tilde{\Phi}(s)\tilde{f}(s)\}\mathcal{L}^{-1}\{s\tilde{\Psi}(s)\tilde{f}(s)\}]$.

To estimate the term $A$, we assume that the number $N$ of bath particles is very large, the internal frequencies can change from 0 to infinity, and replace the sums by integrals according to the rule $\sum_i F(\omega_i) \to \int F(\omega)h(\omega)d\omega$ [14]. The distribution $h(\omega)$ that corresponds to the used memory function is $h(\omega) = (2\xi\omega^2/\pi\mu\tau^2 c^2)/(\omega^2 + \tau^{-2})$. The part of $\tilde{f}_x(t)$ that correlates with $x(0)$ is $f_x(t) = -\mu c^2 x(0)\sum_{i=1}^N \omega_i^{-2}\cos(\omega_i t) \approx -\mu c^2 x(0)\int_0^\infty h(\omega)\cos(\omega t)d\omega = -\xi e^{-t/\tau}x(0)/\tau$. Thus, $A = 2\langle x^2(0)\rangle\mathcal{L}^{-1}\{s\tilde{\Psi}(s)/(s+1/\tau)\}$ at $t \to \infty$ equals zero since $s\tilde{\Psi}(s)/(s+1/\tau) \to QB\tau/(Q^2B^2+\xi^2)$ at $s \to 0$. The term $B$ also tends to 0 at $t \to \infty$. This can be verified by expanding the functions $\tilde{\Phi}(s)$ and $\tilde{\Psi}(s)$ in small $s$ and then calculating the inverse LT in $B$ [20]. The only term that determines the angular momentum at $t \to \infty$ is $C$. Our problem thus reduced to finding the limit of the mean value of

$$L(t) = 2[\mathcal{L}^{-1}\{s\tilde{\Phi}(s)\tilde{f}(s)\}\mathcal{L}^{-1}\{\tilde{\Psi}(s)\tilde{f}(s)\} - \mathcal{L}^{-1}\{\tilde{\Phi}(s)\tilde{f}(s)\}\mathcal{L}^{-1}\{s\tilde{\Psi}(s)\tilde{f}(s)\}]. \tag{7}$$

With the use that the initial values of the functions $\Phi(t)$ and $\Psi(t)$ are zero so that $\mathcal{L}\{\Phi'(t)\} = s\tilde{\Phi}(s)$ and $\mathcal{L}\{\Psi'(t)\} = s\tilde{\Psi}(s)$, the convolution theorem [19], and the FDT, the searched $\langle L(t)\rangle$ is given by to formula

$$\langle L(t)\rangle = \frac{2k_B T\xi}{\tau}\int_0^t dt'\int_0^t dt''[\Psi(t')\Phi'(t'') - \Psi'(t')\Phi(t'')]\exp\frac{-|t'-t''|}{\tau}. \tag{8}$$

Here we have used the time correlation function of the random force,

$$\langle f(t')f(t'')\rangle = \frac{\xi k_B T}{\tau}\exp\frac{-|t'-t''|}{\tau} \tag{9}$$

*Remark.* We will derive the formulas for positive values $m$, $Q$, $B$, $\xi$, and $\tau$. It is evident that changing the sign of the value $Q$ will change the sign of the functions $\tilde{\Psi}(s)$, $\Psi(t)$, and also $\Psi'(t)$. This leads to the change of the sign in the integral (8), and the sign change of the mean value $\langle L(t)\rangle$ and its limit as $t \to \infty$.

III. INVERSE LAPLACE TRANSFORM OF GIVEN FUNCTIONS

It is possible to show that the denominators in $\tilde{\Phi}(s)$ and $\tilde{\Psi}(s)$ have 5 distinct (simple) roots. Denoting

$$\Sigma = \sqrt{(QB\tau)^2 + m(\sqrt{m} - 2\sqrt{\xi\tau})^2} \times \sqrt{(QB\tau)^2 + m(\sqrt{m} + 2\sqrt{\xi\tau})^2}, \tag{10}$$

$$\Pi = m^2 - 4m\xi\tau - (QB\tau)^2, \tag{11}$$

$$\Sigma_+ = \sqrt{\Sigma + \Pi}, \qquad \Sigma_- = \sqrt{\Sigma - \Pi}, \tag{12}$$

we get the roots:

$$s_{1,2} = \frac{-\sqrt{2}m \pm \Sigma_+}{2\sqrt{2}m\tau} + i\frac{\sqrt{2}QB\tau \pm \Sigma_-}{2\sqrt{2}m\tau}. \tag{13}$$

Further $s_{3,4} = s_{1,2}^*$, thanks to the real denominators coefficients and, finally, $s_5 = 0$.

Knowing the denominators' roots, we can determine the coefficients of the partial fraction decomposition in a straightforward way by the substitution of roots after subtracting the root factors, or, formally

$$a_i = \left[(s-s_i)\tilde{\Phi}(s)\right]\big|_{s=s_i}, \qquad b_i = \left[(s-s_i)\tilde{\Psi}(s)\right]\big|_{s=s_i} \tag{14}$$

$i = 1, 2$. We have also $a_{3,4} = a_{1,2}^*$, $b_{3,4} = b_{1,2}^*$, and

$$a_5 = \frac{\xi}{(QB)^2 + \xi^2}, \qquad b_5 = \frac{QB}{(QB)^2 + \xi^2}. \tag{15}$$

Now we can write the wanted function in the complex form

$$\Phi(t) = a_1 \cdot e^{s_1 t} + a_2 \cdot e^{s_2 t} + a_1^* \cdot e^{s_1^* t} + a_2^* \cdot e^{s_2^* t} + a_5. \tag{16}$$

It is possible to show that $b_{1,2} = -ia_{1,2}$. Hence

$$\Psi(t) = -i \cdot a_1 \cdot e^{s_1 t} - i \cdot a_2 \cdot e^{s_2 t} + i \cdot a_1^* \cdot e^{s_1^* t} + i \cdot a_2^* \cdot e^{s_2^* t} + b_5. \tag{17}$$

It is also possible to write functions $\Phi(t)$ and $\Psi(t)$ in real form, however, computation of the integrals in (8) is much simpler in the complex form.

IV. CALCULATION OF THE MEAN VALUE AND ITS LIMIT

So, to calculate the mean value $\langle L(t) \rangle$ in Eq. (8), we deal with next integrals:

$$I_1(c) = \int_0^t dt' \int_0^t dt'' \exp[c(t-t')] \exp\frac{-|t'-t''|}{\tau},$$

$$I_1(c) = \int_0^t dt' \int_0^t dt'' \exp[c(t-t')] \exp\frac{-|t'-t''|}{\tau}$$

and

$$I_{12}(c_1, c_2) = \int_0^t \int_0^t dt' dt'' \exp[c_1(t-t')] \exp[c_2(t-t'')] \exp\frac{-|t'-t''|}{\tau},$$

where $c$, $c_1$, and $c_2$ are complex constants. Indeed, we are first of all interested in the limit $\lim_{t\to\infty}\langle L(t)\rangle$. To find a formula for such a limit, it is sufficient to use the formulas

$$\lim_{t\to\infty} I_1(c) = \lim_{t\to\infty} I_2(c) = -\frac{\tau(\tau c - 2)}{c(\tau c - 1)} \tag{18}$$

and

$$\lim_{t\to\infty} I_{12}(c_1, c_2) = \frac{\tau[\tau(c_1 + c_2) - 2]}{(c_1 + c_2)(\tau c_1 - 1)(\tau c_2 - 1)}, \tag{19}$$

which are derived using the conditions $\tau > 0$, $\text{Re}(s_k) < 0$, $k = 1, 2, 3$, and 4.

The integrals $I_1(c)$, $I_2(c)$, and $I_{12}(c_1, c_2)$, could be calculated, but the general formula for $\langle L(t)\rangle$ is too complicated, and indeed it is not necessary to get it. Mean values $\langle L(t)\rangle$ could be evaluated for given values $t$ using a computer algebra system. We have used CAS Maple.

Although the formula for the limit $\lim_{t\to\infty}\langle L(t)\rangle$ is much simpler than the general formula of $\langle L(t)\rangle$, we were unable to find it using Eqs. (16) and (17) together with (18) and (19). Fortunately, by a happy coincidence, we managed to guess the formula, which is as follows:

$$\lim_{t\to\infty}\langle L(t)\rangle = -2\frac{k_B T}{\xi}\frac{QB/\xi}{(QB/\xi)^2 + 1}, \tag{20}$$

In particular, for $B \to 0$ we get $\lim_{t\to\infty}\langle L(t)\rangle \approx -2k_B T Q B / \xi^2$.

Figure 1 shows the relative value $\langle L(t)\rangle$ divided by the $\lim_{t\to\infty}\langle L(t)\rangle$ as a function of $t/\tau$ for $Q = 1.6 \cdot 10^{-19}$ C and the rest of the fixed parameters are estimated as for a methane molecule in water at room temperature, $\xi = 2.0 \cdot 10^{-12}$ kg/s, $\tau = 4.0 \cdot 10^{-14}$ s, and $m = 3.0 \cdot 10^{-26}$. Fig. 2 shows that the mean value $\langle L(t)\rangle$ converges to the limit very fast in an exponentially oscillattory manner. The relative difference is defined as $\varepsilon(t) = (\langle L(t)\rangle - \lim_{t\to\infty}\langle L(t)\rangle) / \lim_{t\to\infty}\langle L(t)\rangle$.

## V. CONCLUSION

In conclusion, we have considered the BM of a charged particle in a neutral bath when this system is placed in a constant magnetic field. The ZCL model that is frequently and effectively used to describe the particle dynamics with memory is found to give a nonzero averaged diamagnetic moment at long times when the system is expected to be in equilibrium. This finding is surprising if one expects that the BvL theorem holds for the BP itself. However, our result does not necessarily contradict the BvL theorem when applied to the open system BP in a bath, where the bath constantly interacting with the BP can be subtly driven out of equilibrium by the particle's presence in a magnetic field [11]. Another interesting problem is connected with our counter-intuitive finding that according to the simple formula proposed by us, the angular momentum (and thus magnetic moment) of the BP goes to zero as the magnetic field $B$ approaches infinity. This might occur if the specific damping in the ZCL model becomes so effective that beyond a certain magnetic field strength, it suppresses the net angular momentum. It implies that the energy imparted by the magnetic field to induce angular motion is increasingly dissipated by the bath at higher field strengths, leading to a net zero angular momentum. Another possibility of explaining such an effect, mentioned in [10], is related to the Larmor radius of the circular path of a charged particle in a magnetic field $B$, which becomes infinitesimally small as $B$ becomes extremely strong. We also refer to the work [10] for a discussion of a demanding problem of experimental realization of the studied classical system.

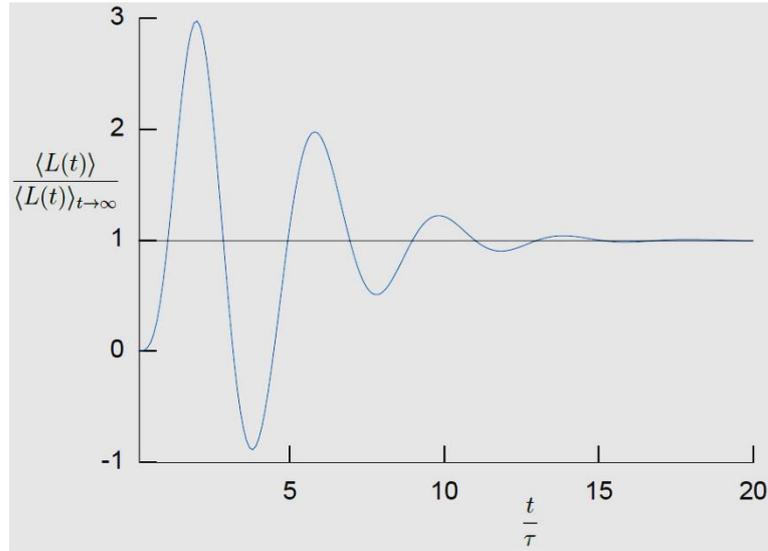

**Fig. 1.** Mean value $\langle L(t) \rangle$ divided by the $\lim_{t \to \infty} \langle L(t) \rangle$ as a function of $t/\tau$.

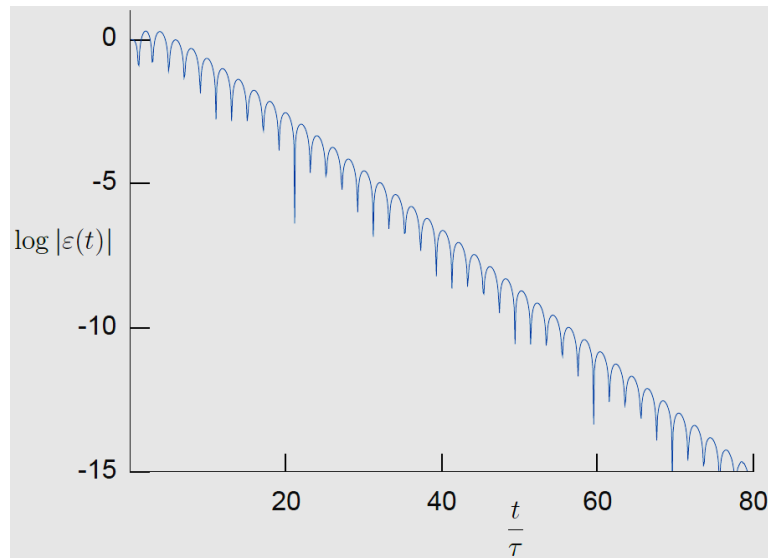

**Fig. 2.** Logarithm of the absolute value of the relative difference between $\langle L(t) \rangle$ and $\lim_{t \to \infty} \langle L(t) \rangle$.

ACKNOWLEDGMENT

This work was supported by the Scientific Grant Agency of the Slovak Republic through the grant VEGA 1/0353/22. All Maple calculations were performed using the MLIT License Management System.